\documentstyle[12pt,aasms4,epsfig]{article}

\begin{document}

\title{A 3 dimensional  diagnostic diagram for Seyfert 2s: 
probing X-ray absorption and Compton thickness}

\author{L. Bassani\altaffilmark{1}, M. Dadina\altaffilmark{1}}
\affil{Istituto Te.S.R.E./CNR, Via Gobetti 101, I-40129 Bologna, Italy}

\author{R.Maiolino\altaffilmark{2}, M.Salvati\altaffilmark{2}}

\affil{Osservatorio Astrofisico di Arcetri, L.E. Fermi 5, I-5015 Firenze, 
Italy}

\author{G. Risaliti\altaffilmark{3}}
\affil{Dipartimento di Astronomia, Universita' di Firenze,L.E. Fermi 5, 
I-5015 Firenze, Italy}

\author{R. Della Ceca\altaffilmark{4}}
\affil{Osservatorio Astronomico di Brera, Via Brera 28, I-20121 Milano,
Italy}

\author{G. Matt\altaffilmark{5}}
\affil{Dipartimento di Fisica, Universita' di Roma III, Via della Vasca
Navale
84, I00146 Roma, Italy}

\author{G. Zamorani\altaffilmark{6,7}}
\affil{Osservatorio Astronomico di Bologna, Via Zamboni 33, I-40126
Bologna, Italy}
\affil{Istituto di Radioastronomia/CNR, Via Gobetti 101, I-40129 
Bologna, Italy}

\begin{abstract}
We present and discuss a "3-dimensional" diagnostic diagram for
Seyfert2 galaxies obtained by means of X-ray and [OIII] data on 
a large sample of objects (reported in the Appendix). 
The diagram shows the K$\alpha$ iron line
equivalent width as a function of both  the column density derived from
the photoelectric cutoff and the 
2-10 keV flux normalized to the [OIII] optical line flux (the latter corrected
for extinction and assumed to be a true indicator of the source intrinsic
luminosity). 
We find that  the hard X-ray properties of type 2 objects depend
on a single parameter, the absorbing column density along the line of sight,
in accordance with the unified model.  The diagram can be used to identify
Compton thick sources and to  isolate and study peculiar objects.
From this analysis we have  obtained a column density distribution 
of Seyfert 2 galaxies which is thought to be a good approximation  
of  the real  distribution. A  large population  of heavily absorbed
objects is discovered, including many Compton thick candidates. Our results
indicate that the mean Log N$_{\rm H}$/ cm$^{-2}$   in type 2 Seyferts is 
23.5 and that as much as 23-30$\%$ of sources have 
N$_{\rm H}$ $\ge$ 10$^{24}$  cm$^{-2}$.

\end{abstract}

{\em Subject headings}: Active galactic nuclei- Seyfert 2 galaxies-
Seyfert 2-high energy spectra; X-rays- [OIII] line emission

\section{Introduction}

The discovery of broad emission lines in the
polarized spectrum of a number  of Seyfert 2 galaxies,
has provided the first observational evidence in favour of the unified model 
which states that the main discriminating parameter between the two Seyfert
(Sey) types is the inclination of our line of sight with respect to an 
obscuring torus
surrounding the nucleus (see Antonucci 1993 for a review).
X-ray data have subsequently reinforced this idea  by 
demonstrating  that type 2 objects are generally characterized by
strong absorption, corresponding to column densities N$_{\rm H}$ $\ge$ 
10$^{22}$ cm$^{-2}$ (Smith $\&$ Done 1996, Turner et al. 1997a,
Maiolino et al. 1998). 
X-ray measurements of the column densities in Sey2s
proved to be important  in our understanding the nature 
of the absorbing medium, the validity of the unifying theory and also
the relevance of type 2 objects to the synthesis of the X-ray 
background. \\
The intrinsic absorption also defines the source characteristics:
for N$_{\rm H}$$\le$ 10$^{24}$ cm$^{-2}$, X-rays
above a few keV can penetrate the torus making the source nucleus visible to
the observer and the column density measurable; in this case the source 
is called "Compton thin". 
For values of N$_{\rm H}$ around a few 10$^{24}$~cm$^{-2}$, only X-rays 
in the 10-100 keV range pass through the torus and so only data in this 
band allow an estimate of the source column density.
For values of N$_{\rm H}$ higher than 10$^{25}$~cm$^{-2}$, also 
X-rays above a few tens of keV
are absorbed (as photons, after a few scattering, 
are redshifted down to the photo-absorption regime) and  the nucleus 
is totally hidden to our view; in these sources only emission 
reflected by the torus
(cold reflected component) or scattered by warm material near the nucleus 
(warm scattered component) is observed. In this case, the photoelectric cut-off 
(if any) in the observed spectrum does not provide information 
on the real column density 
absorbing the primary X-ray source and so the galaxy may be  erroneously 
classified  as a low absorption object; therefore, for N$_{\rm H}$ $\gg$
10$^{24}$ cm$^{-2}$, the X-ray data can only provide a lower limit to the 
value of the column density. These so called "Compton thick" 
sources are extremely faint 
in X-rays and therefore until recently only a few were known
(Matt 1997).
The improvement in sensitivity of current X-ray telescopes has allowed  
few more sources of this type to be  discovered  
(Turner et al. 1997b, Maiolino et al. 1998). 
 
Evidence for the presence of obscuring material can also be obtained
via X-ray spectroscopy as iron line emission is  expected to be 
produced either via
transmission through (Leahy $\&$ Creighton 1993) or scattering/reflection 
by (Ghisellini et al. 1994, Matt et al. 1996a) the absorbing material . 
In most Compton thin cases
the observed equivalent widths (EW) are of the order of a few hundred eV, 
consistent with the typical spectra of Sey1s transmitted through the
observed column densities (Turner et al. 1998).
When the column density increases to a few 10$^{23}$ cm$^{-2}$, the line EW 
increases as it is measured against a depressed continuum; the
EW can be higher than 1 keV for column densities  $\ge$ 
10$^{24}$  cm$^{-2}$ and
such values are indeed  observed in highly absorbed objects and 
Compton thick sources (Matt 1997, Maiolino et al. 1998).
Although it is tempting to identify all sources with high equivalent width 
as Compton thick systems, such condition
can also occur if the ionizing radiation is anisotropic 
(Ghisellini et al 1991), 
or if there is a lag between a drop in the continuum and in the 
line emission as 
indeed observed in NGC2992 by  Weaver et al. (1996).
Therefore when looking for Compton thick candidates, it is important
to take other evidence into consideration.

Measuring the X-ray luminosity against an isotropic indicator of the 
intrinsic brightness of the source offers an alternative method for evaluating
N$_{\rm H}$: assuming that the 
unifying theory is correct, 
the X-ray flux is depressed with respect to this isotropic indicator 
by an amount related to the absorbing column density. Many authors
have used the [OIII]$\lambda$5007 and  Far-infrared (hereafter [OIII] and FIR
respectively) luminosities as isotropic
indicators of the source nuclear strength, without however considering their
limitations: the [OIII] flux may be affected by large scale obscuration in
the host galaxy (Maiolino $\&$ Rieke 1995, Hes et al. 1993), while the 
FIR flux  may be
contaminated by a starburst component (Maiolino et al. 1995). While it is 
difficult to account  for the starburst component, it is possible 
to correct the  [OIII] emission for the extinction towards the 
Narrow Line Region as deduced from the 
Balmer decrement.
Studies performed in the past have found  that the ratio between the 
2-10 keV X-ray and [OIII] fluxes (hereafter T or thickness parameter) cover 
the range 1-100 for both type 1 and type 2 objects. This was taken as 
evidence that the torus column density is never 
high on average (Mulchaey et al. 1994, Alonso-Herrero et al. 1997). 
However, these results were
strongly biased for type 2 objects as it will become apparent in the following.

\section{A 3D diagnostic diagram in the X-ray band}

As the X-ray band provides 3 ways to measure the absorption
along the line of sight (the photoelectric
cutoff, the iron line EW and the thickness
parameter T) it is convenient to plot Seyfert galaxies in this
3-dimensional parameter space.

Figure 1  is  a plot of T versus EW for a large sample of Sey2s, 
with  the value of the column density derived from the photoelectric cutoff
coded in the symbols as indicated.
In the Appendix, we present the database used for the present study.
It consists of a sample of 73 objects 
for which "good" X-ray spectra were available in 
the literature (see the Appendix for details on the selection criterion). 
The spectral fit of the hard X-ray data of these sources were quite
homogeneous: an absorbed power law and a gaussian to account for the Fe 
line. [OIII] fluxes were also collected from the literature.
Particular care has been taken to correct the [OIII] fluxes 
for reddening affecting the Narrow Line Region; this correction 
is not negligible in  several cases (see Appendix) and provides a more 
meaningful estimate of the T parameter with respect to previous studies
(Mulchaey et al. 1994, Alonso-Herrero et al. 1997).
A few objects, although present in the Tables of the Appendix,
have  been discarded from the following analysis because of dubious data:
F09104+4109 and Cygnus A have X-ray data probably contaminated by 
intracluster emission while for  NGC5128 it is difficult to estimate
both the [OIII]
flux (see Appendix) and the nuclear X-ray flux (Turner et al. 1997c).
In three cases (NGC3031/M81, NG6251 and IRAS20210+1121) a 6.7 keV line is the 
only iron K${\alpha}$ line detected and so its EW is plotted in the 
figure.

\begin{figure}
\centerline{\vbox{
\epsfig{figure=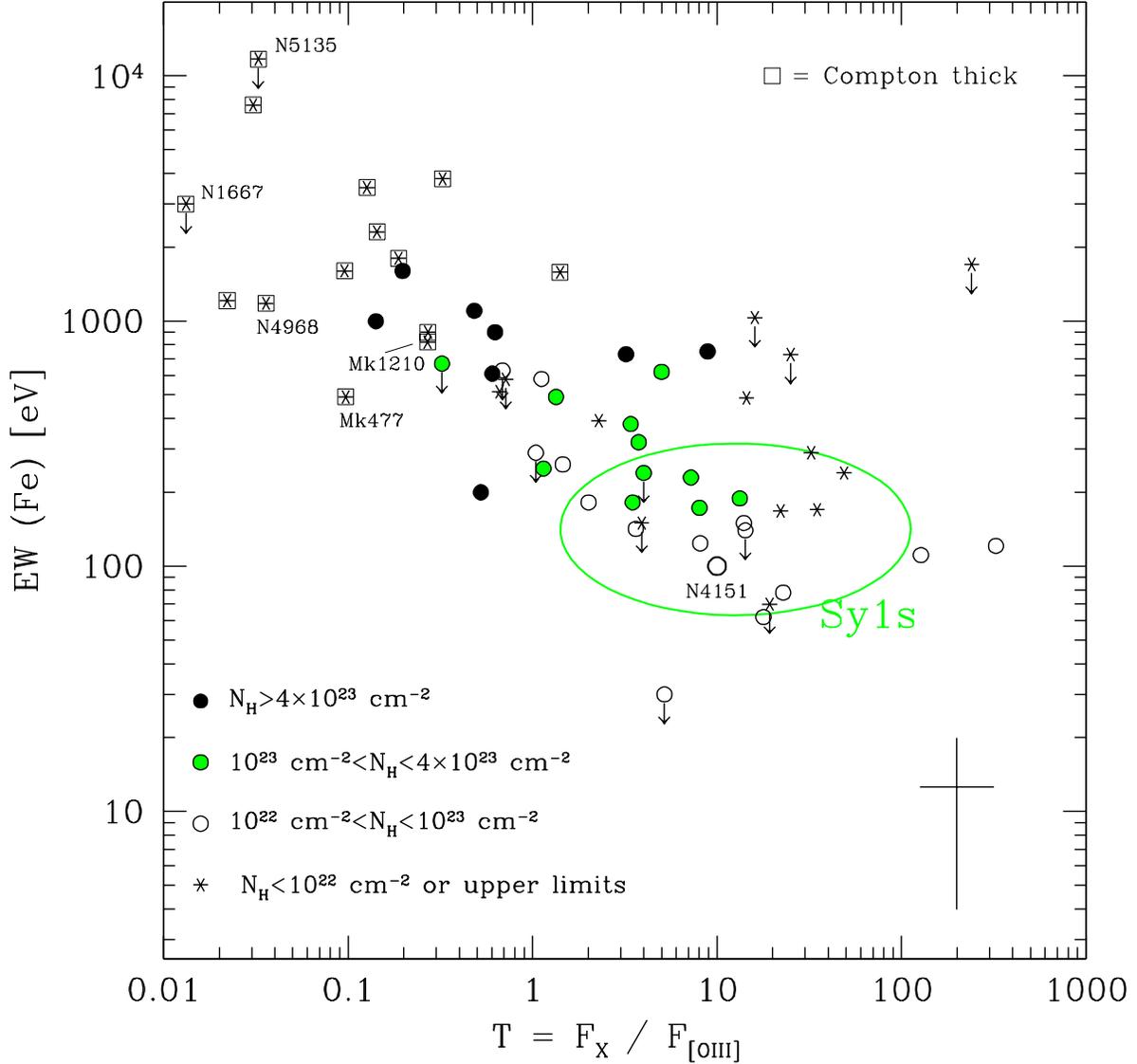, height=16cm, width=16cm, angle=0}
}}
\caption{The distribution of T=F$_{\rm X}$/F$_{\rm [OIII]}$ 
as a function of the iron K$\alpha$ line EW, with the
value of N$_{\rm H}$ (derived from the photoelectric cutoff) 
coded by the indicated symbols. The Compton thick sources have
a square around the symbol corresponding to the  N$_{\rm H}$
value.  The cross
indicates the typical error associated to each measurement, while the ellipse
represents the region where the Sey1s in the sample of Mulchaey et al (1994)
are found. Sources discussed in the text are labelled in the diagram, which 
also show the position of the intermediate Sey galaxy NGC4151.
}
\label{}
\end{figure}  
        
An  immediate result of figure 1 is a T distribution quite different from 
previous reports; in particular T can be $\ll$ 1, in a significant fraction
of Sey2s whereas all well studied
Sey1s have T$\ge$ 1 (Maiolino et al. 1998).
Since the sample used in building up this diagnostic diagram is highly
heterogeneous, the relative populations of the various portions of
it may not correspond to the real distribution. However the 
general shape of the diagram, which appears to be qualitatively in
agreement with the unified theory and with the assumption that
the dereddened [OIII] flux is an isotropic indicator of the
intrinsic luminosity of the Sey nucleus, is meaningful and  translates 
into quantitative 
terms the relations between the 3 absorption indicators which were
sketched qualitatively in the introduction. 

The relationship shown in figure 1 can be used to confirm
already known Compton thick sources and to suggest new candidates;
in particular, sources where the flux is too weak for spectral fitting
and T is known to be $\le$ 0.1 are likely to be highly absorbed objects
(i.e. with N$_{\rm H}$ $\ge$ 10$^{24}$ cm$^{-2}$). 

In  figure 1, 14 likely Compton thick sources are shown (see also Table 2
of the Appendix).
These have a square around the symbol corresponding to the  N$_{\rm H}$ value.
Most of them have already been proposed as  Compton thick sources in the 
literature (see for example Matt (1997) for NGC1068, Circinus galaxy and 
NGC6240, Malaguti et al.(1998) for  NGC7674, Ueno et al. (1998) for F20210+1121 
and Maiolino et al. (1998) for NGC1386, NGC2273, NGC3393 and NGC5643) 
and we confirm this classification on the basis of our diagram.  
Note that since for the sources discussed by Maiolino et al. (1998) the X-ray
parameters are deduced from a simple model (see Appendix), NGC4939 is
classified here as Compton thin although its nature is still uncertain.

\begin{figure}
\centerline{\vbox{
\epsfig{figure=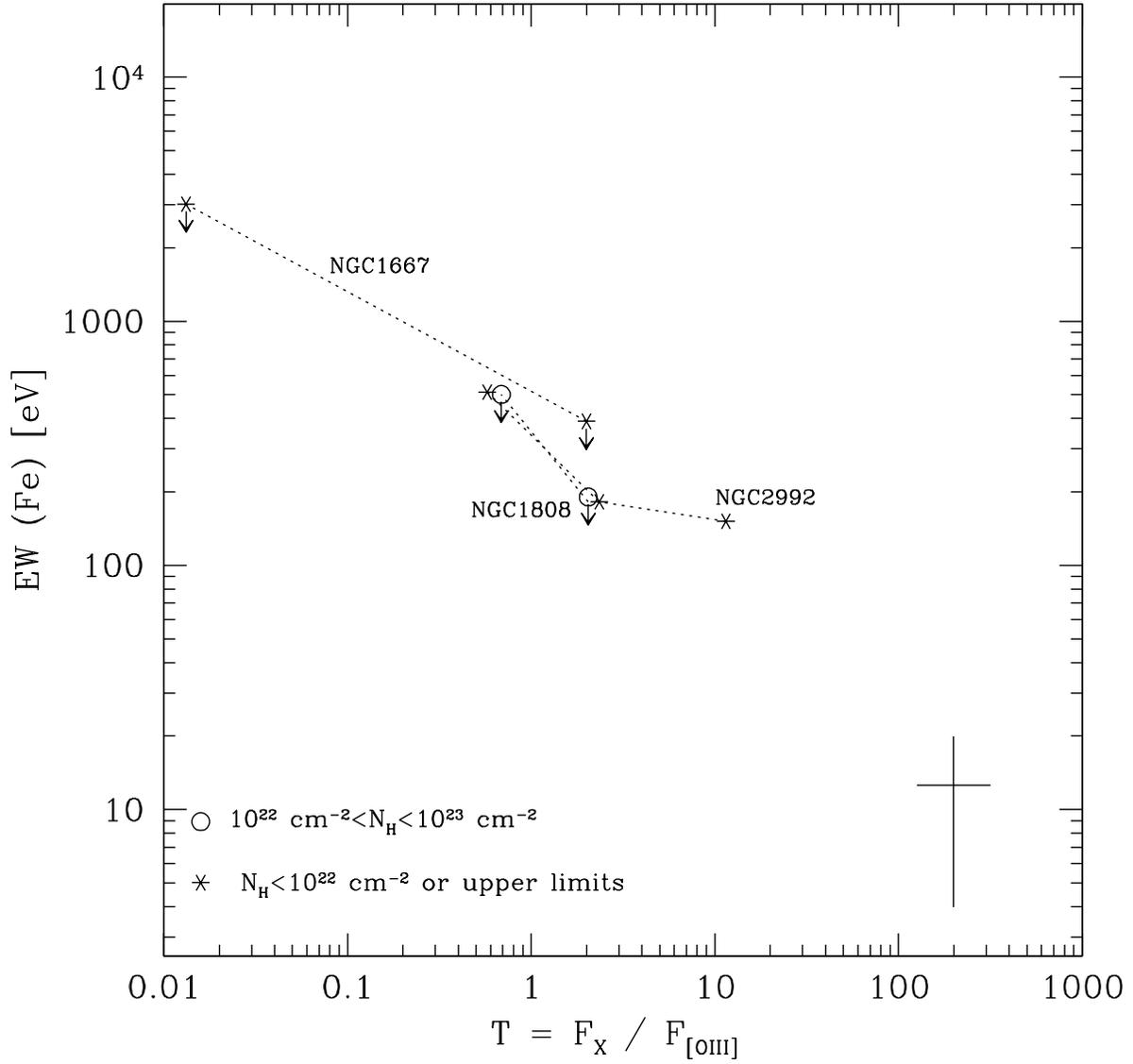, height=16cm, width=16cm, angle=0}
}}
\caption{As in Figure 1 but only for "dying" AGNs; the 
temporal evolution of their location in the diagram is shown by dotted lines}. 
\label{}
\end{figure}
The same diagram suggests five  new Compton thick candidates: NGC5135, NGC4968
NGC1667, MKN1210 and MKN477.
The cases of NGC5135 and NGC4968 were  considered inconclusive by Turner et al.
(1997b) because of
the large uncertainty on the line EW. However  the low value of T 
for both of them, 
almost two orders of magnitude lower than the average in  Sey1s,  and the 
small measured column
density, consistent with being entirely due to galactic absorption, 
make them very likely Compton thick candidates. The same
considerations apply to NGC1667 whose classification was also 
considered uncertain by Turner et al.
Note, however, that an alternative interpretation for the location 
of this  source in the diagnostic diagram is possible (see discussion below).
MKN1210 and MKN477 can  also be considered as Compton thick objects according
to our diagnostic diagram, although their nature is less well defined than in
previous cases.
In particular, the EW is lower than expected in MKN477
but both the low T value and the observed flat X-ray spectrum, typical of
reflection dominated sources, are consistent with our suggestion. 
On the other hand, the  Compton thick nature of three other sources  
discussed by Turner et al. (1997b), namely  MKN 273, MKN463 and NGC2992 is
still questionable according to figure 1. In MKN 273
and MKN463, the 6.4 keV line EW 
is smaller than  $\sim$ 600 keV and compatible with the observed column 
densities while the not so low values  of T,  0.7 and 0.3 respectively,  
are consistent with 
Sey1 values if a correction for absorption is applied to the X-ray flux. 
Different is the case
of NGC2992: Weaver et al. (1996) studied this source and reported a systematic
decrease in the 2-10 kev luminosity by a factor of $\sim$16 over the last 
20 years, while the iron line flux decreased only by a factor 2-3 and the 
reflection component became 5 times stronger.   These authors interpreted 
the data in terms of a lag between reprocessed and intrinsic fluxes and 
accordingly located the reprocessor at the torus. We rephrase their statement
and suggest  that when an active nucleus fades away it moves along
the diagnostic diagram at a constant column density, in a way which depends
on the geometry and physics of the reflecting material.
The temporal "evolution" of NGC2992 in our diagnostic diagram is shown in 
figure 2, based on data in Table 1-2 of the Appendix and 
past data (Weaver et al. 1996). 
Another object similar to NGC2992 is NGC1808,  
which declined significantly in flux 
from Ginga to ASCA observation (Polletta et al. 1996, 
Turner et al. 1997a), thus moving in the diagram towards the region of
high EW and low T values (see figure 2). It is therefore possible that, at a 
certain stage of their fading process,  these "dying"
objects can be easily mistaken for Compton thick sources. This ambiguity
is for example present in the case of  NGC1667 (also displayed in figure 2), 
which historically has varied
by a factor of 150 in the 2-10 keV band (Polletta et al. 1996, Turner et al. 
1997a) moving in T from 1.5 to 0.01; only future observations 
will be able to assess the correct interpretation for this object. 
For the time being,
we consider NGC1667 as a likely Compton thick 
candidate  also in view of the significant uncertainties associated 
with the Ginga observation (Smith and Done 1996).
 
\section{The distribution of absorption column densities in Sey2s}

A 3D plot like figure 1 for  a sample with known selection and 
completeness properties would provide statistically significant  
information on  the true distribution of 
absorption column densities. Such an information would be extremely valuable 
for constraining the geometric parameters of the torus. The available sample 
is not statistically valid,
nevertheless it provides some bounds to this true distribution.
Figure 3 (upper panel) reports the distribution of column densities 
for all sources in our sample excluding objects with upper limits on
N$_{\rm H}$ , but including MCG-05-18-002, which has been 
classified as a Compton thick source by Maiolino et al. (1998) simply
on the basis of its low T value. 
Sources identified as Compton thick, i.e. with 
N$_{\rm H}$ $\ge$ 10$^{24}$ cm$^{-2}$ are indicated  by a shaded area 
in the histogram;  in particular since only data above 10 keV can rule out
an absorbing column density in the range
10$^{24}$-10$^{25}$ cm $^{-2}$, Compton thick sources for which these data are 
lacking are placed at N$_{\rm H}$ $\ge$ 10$^{24}$ cm$^{-2}$, while the others 
are placed at N$_{\rm H}$ $\ge$ 10$^{25}$ cm$^{-2}$.
The most remarkable result of our analysis is the discovery of a
large population of highly absorbed sources:
more than  half of the sample has N$_{\rm H}$ $\ge$ 10$^{23}$ cm$^{-2}$, with 
25$\%$ of the sources having N$_{\rm H}$ $\ge$ 10$^{24}$ cm$^{-2}$.

\begin{figure}
\centerline{\vbox{
\epsfig{figure=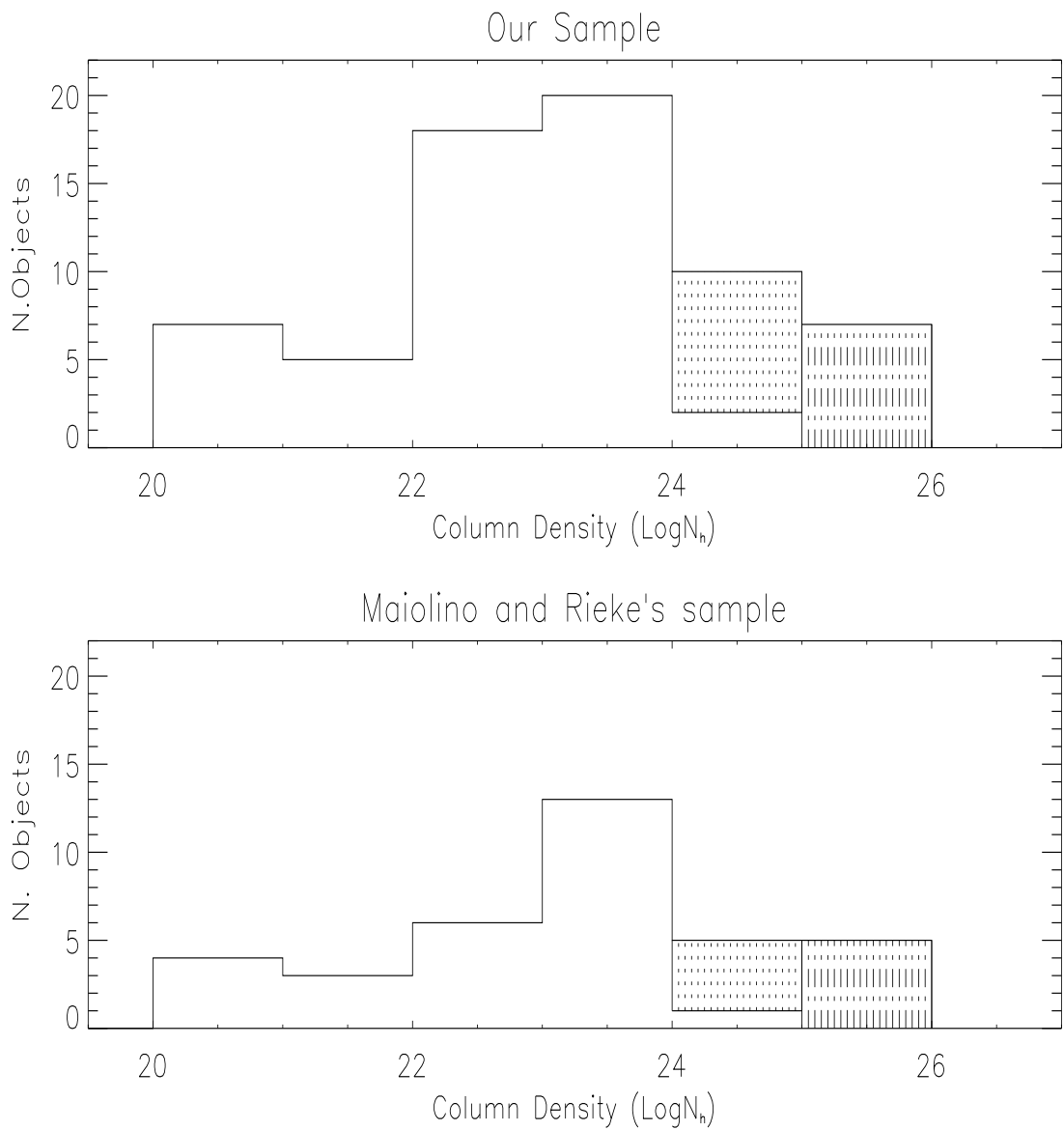, height=16cm, width=16cm, angle=0}
}}
\caption{Distribution of the absorbing column densities  for our
X-ray  sample (upper panel) and  for the subsample
of objects contained in the Maiolino \& Rieke's  catalog
(lower panel). The shaded area bins contain objects for which only a lower 
limit on N$_{\rm H}$ is available; in particular the first
of such bin refer to objects identified as 
Compton thick with N$_{\rm H}$ above 10$^{24}$ cm$^{-2}$ for lack of
high energy X-ray data. Five objects (IC1631, NGC1672, MKN273,
AX1749+684 and IC1368), for which only an upper limit on the column density
is available,  are not included in the upper panel histogram}
\label{}
\end{figure}  

In order to proceed beyond this statement, we have considered the 
intersection between  our X-ray sample and the sample of Sey2s defined by
Maiolino $\&$ Rieke (1995) .
The latter is an optical spectroscopic sample drawn from the 
RSA catalog of galaxies, which in turn is magnitude limited with respect
to the host galaxy. According to the extensive discussion presented by the 
authors, the Maiolino $\&$ Rieke sample of local active nuclei is much
less affected by orientation and luminosity biases than any other similar 
sample.
Out of the  71 Sey2s contained in Maiolino $\&$ Rieke (1995), 36 are in our
database and their column density histogram is also shown in figure 3 (lower
panel). For this sample $\sim$ 64$\%$ of the objects has  
N$_{\rm H}$$\ge$ 10$^{23}$ cm$^{-2}$ (even more if intermediate types are 
excluded) while the  28$\%$ with N$_{\rm H}$$\ge$ 10$^{24}$ cm$^{-2}$
correspond to Compton thick or nearly Compton thick objects. 
The distribution has a mean Log N$_{\rm H}$/cm$^{-2}$ 
of $\sim$  23.5, significantly higher
than what determined in the past.

We have also considered the intersection of our database and the extended
12 ${\micron}$ sample (Rush, Malkan $\&$ Spinoglio 1993), which is a flux 
limited sample of 893 galaxies selected according to infrared properties.
In this case, 31 objects appear in our X-ray catalog (irrespective of their
classification in the 12 ${\micron}$ sample) and their N$_{\rm H}$ distribution
is similar to those reported in figure 2: half of the sample (52$\%$)
has N$_{\rm H}$ $\ge$ 10$^{23}$ cm$^{-2}$, with 23$\%$ of the sources having
N$_{\rm H}$ $\ge$ 10$^{24}$ cm$^{-2}$. The results on both these sub-samples
are in line with the finding of
Maiolino et al. (1998) that the average obscuration of Sey2s, selected 
solely on the basis of their [OIII] flux, is much higher than found previously.

However, also the  N$_{\rm H}$ distribution of these sub-samples may still be
affected by 
biases. Even if the optical/infrared selection is unbiased, 
the X-ray selection
(expecially for the EXOSAT, Ginga and the oldest ASCA data) favors
sources already detected in previous all-sky surveys and therefore
X-ray bright and/or little absorbed. We can then conclude that the
true  N$_{\rm H}$ distribution is likely to be even more shifted towards high 
N$_{\rm H}$ values. 
An estimate of the effect of this residual bias (hence an estimate 
of the "true"  N$_{\rm H}$ distribution) is presented in 
Risaliti (1998), who identifies the dominant bias in the various
subsamples contributing to the X-ray database and extrapolates the
partial N$_{\rm H}$ distributions to the entire Maiolino $\&$ Rieke sample.
At any rate,  the large fraction of heavily obscured Sey2s that are 
beginning to be found
may have  important implications for the synthesis of the X-ray
background and has to be taken in due account in future refinements
of existing models (see for example Comastri et al. 1995).

Finally, note that a few sources in figure 3 have N$_{\rm H}$ values
consistent  with pure 
galactic absorption (i.e $\sim$ a few 10$^{20}$ cm$^{-2}$); they also have
T values typical of Sey1s. Some of them 
(M81 and NGC 5033) may be wrongly classified as Sey2s
and probably are intermediate type 1.5 objects; also NGC4579 is 
probably a LINER, so that the low
N$_{\rm H}$ value is compatible with its classification (see Appendix). 
On the other hand,
NGC7590 and NGC3147 are bona-fide Sey2s in the optical and yet have 
negligible absorption.
The low N$_{\rm H}$  could be reconciled with 
their Sey2s nature only if they were Compton thick, but this is not supported
by their large T values (7 and 14, respectively) .
Also, NGC3147 has  an iron line equivalent  width of 
500$\pm300$ eV (consistent  with a high absorption),
but no line is detected in NGC7590. It is therefore tempting to 
conclude that there might be  a few objects in which the Sey2s appearence
is intrinsic and not due to obscuration. If confirmed, this would 
revive the long standing idea of a population of Sey2s 
(the "true Sey2s"?) which just  miss their broad line region
(Tran 1995).  Obviously more sources of this type should be
identified before confirming this hypothesis.

\section{Conclusions}
              
We have presented a "three dimensional" diagnostic diagram in the X-ray band
for a large sample of Sey2s. 
The diagram shows the K$\alpha$ iron line
equivalent width as a function of both the N$_{\rm H}$ derived from the
observed photoelectric cutoff and  the 
2-10 keV flux normalized to the [OIII] line flux (the latter corrected
for extinction in the Narrow Line Region and assumed to be a true indicator 
of the source intrinsic luminosity).
The source distribution in this  3-dimensional parameter space is in 
qualitative agreement 
with the unified model, suggesting that the optical thickness of the torus 
along the line of sight is the main parameter responsible for the source
X-ray properties. The diagram is useful to 
calibrate  N$_{\rm H}$ as a function of the  flux ratio T: for faint sources 
this may be the only way to determine their Compton thick nature. It is also 
useful to isolate and study peculiar objects like "dying" AGNs (like NGC2992)
and "true" Sey2s (like NGC7590). 
Finally, and most importantly,
our results provide an estimate of the true  N$_{\rm H}$ distribution 
for type 2 objects, which is an important input for 
the synthesis models of the X-ray background.
From our sample we find a mean column density 
log N$_{\rm H}$/cm$^{-2}$ $\sim$ 23.5,
significantly larger than what determined in the past. We also find
a large fraction of heavily absorbed objects, including a high
percentage of Compton thick systems;
out of these, several are newly classified ones.

{\bf Acknowledgement}\\

M.D. acknowledges the CNR for a fellowship and the TeSRE Institute for the
hospitality. This research has made use of data obtained through (1)NASA/IPAC
Extragalactic Database (NED) which is operated by the Jet Propulsion
Laboratory,Caltech under contract with
the National Aeronautics and Space Administration, (2)the High Energy
Astrophysics Science Archive Research Center (HEASARC) on line 
service 
provided by NASA-Goddard Space Flight Center and (3) the SIMBAD database
at CDS, Strasbourg, France. Financial support by MURST, CNR and ASI is
gratefully acknowledged;  in particular, M.S., R.M. and G.R. have been 
partially supported by ASI through grant ARS-96-66 and G.Z. 
through grant ARS-96-70.

\clearpage

\section{APPENDIX. The Catalog}

The catalog presented in this Appendix contains all Sey 2s for which a 
'reliable' hard X-ray spectrum was
available in the literature, i.e those galaxies having 
a spectrum observed in the 2-10 keV band that allows the determination of 
at least three of the following quantities: the photon index, the absorbing 
column density derived from the photoelectric cutoff, the 2-10 keV 
flux and the EW of the iron K$\alpha$ line at 6.4 keV. 
In a few objects, the reported iron line energy is
higher than 6.4 keV; these cases are highlighted by a note.
Only objects classified as  Seyferts of 
type 1.8, 1.9 or 2.0 either in 
NED, SIMBAD or according to Maiolino and Rieke (1995), 
have been included in the
catalog. A few exception must 
be justified. The recently discovered galaxies AXJ131501+3141 and 
AXJ1749+684 have been included 
on the basis of their type 2  characteristics (\cite{akiy98,iwas97a}).
For some objects in the sample the classification based on  their optical
spectra is uncertain. In particular, some 
have features intermediate between  those of Seyferts and LINERs,  while 
intense star formation is present in some other  sources. We also came 
across objects with uncertain Seyfert 2 classification 
(for example some sources  
have been recently reclassified as type 1-1.5  objects by \cite{mora96},
\cite{ho97} and 
\cite{iwas98}).
These ambiguous, intermediate or composite  classifications  are indicated 
in Table 1 whenever possible. 

\placetable{tbl-1} 

In order to estimate the thickness parameter T=F$_{\rm X}$/F$_{\rm [OIII]}$, 
the catalog
has been complemented with additional information on  
the [OIII] emission line flux and on the Balmer decrement (the latter
aimed at the extinction correction). 
The  [OIII] fluxes were obtained by various sources 
in the literature where we  favored  apertures larger than $\sim$ 100pc
as projected on the source, so that most of the [OIII] emission coming from 
the Narrow Line Region is included. The contribution to the [OIII] flux from
circumnuclear star forming regions is usually negligible. 
We  favored data with higher signal to noise 
ratio and most recent publications. 
Also, to determine the Balmer decrement, we preferred data that were corrected
for H$\beta$ stellar absorption.
Whenever more than one paper provided data of 
similar quality, the average value for both the Balmer decrement and 
the [OIII]
line flux was estimated and in the [OIII] case also the standard deviation 
between the various data was  reported in the table as an 
uncertainty estimate.
The [OIII] fluxes were corrected for optical reddening whenever
a value of the Balmer decrement was available by using the following relation:
\begin{center}
F$_{[OIII],cor}$= F$_{[OIII],obs}$ $\times$
[(H$\alpha$/H$\beta$)/(H$\alpha$/H$\beta$)$_{0}$]$^{2.94}$
\end{center}
and assuming an intrinsic 
Balmer decrement (H$\alpha$/H$\beta$)$_{0}$ =3. Whenever a broad component for
 H$\alpha$ or H$\beta$ was present (Seyfert 1.8 and 1.9), only the narrow 
components were used. 

The sample presented here contains 73  objects (57 of type 2.0, 11 of 
type 1.9, 4 of type 1.8 and 1 NLXG) which are listed in Table 1 along with  
optical position in equatorial coordinates for epoch J2000 (col.[2]), 
redshift z (col. [3]) as reported in NED (NASA Extragalactic Database), 
Seyfert type (col.[4]), galactic HI column density from 21 cm measurements
in units of 10$^{20}$ cm$^{-2}$ (col.[5]) from HEASARC (High Energy 
Astrophyisics Science Archive Research Center)  on line service,
 Balmer decrement (col.[6])
and reddening corrected [OIII] emission line flux (col.[7]) in units of 
10$^{-11}$
erg cm$^{-2}$ s$^{-1}$. References relative to the Balmer decrement as 
well as to the [OIII] flux are quoted as small  numbers 
in col. [6] and [7] respectively and reported at the end of the Table.
The origin of the classification is reported in small
letters next to col. [4]. The redshift of most sources falls in the
interval 0.002$<$ z $<$0.03
with the largest redshift in the sample being z$\sim$0.4.
 
X-ray data are  presented in Table 2, which reports 
the photon index $\Gamma$ (col.[2]),the  hydrogen column density N$_{H}$ 
in units of 10$^{20}$ cm$^{-2}$ 
derived from the photoelectric cutoff  (col. [3]), the fluorescence  
iron line Equivalent Width in eV (col.[4]) as well as the 2-10 keV flux 
both observed (col.[5]) and corrected
for the measured absorption (col.[6]) in units of 10$^{-11}$ 
erg cm $^{-2}$ s$^{-1}$; data reported in col.[2],[3] and [4] are listed 
together with their associated errors (generally these corresponds
to  90$\%$ confidence level except in a few cases, see for example 
\cite{turn97a}). The majority of 
our X-ray data come from either Ginga or ASCA observations 
supplemented with recent measurements performed by the BeppoSAX 
satellite(\cite{maio98,mala98,ueno98}). 
Most  of these hard X-ray data (which are reported here as quoted in 
the original reference) 
were fitted with an absorbed power law and a gaussian to account
for the iron line. Some authors presented also
alternative fits (eg. reflected continua as done 
for example in \cite{turn97a,maio98}); but we always
adopted the transmission model (absorbed
power law), both for homogeneity with the other data and to avoid the
a priori assumption of a Compton thick model (the Compton thick nature 
of some sources
is what we want to determine a posteriori by means of our diagram).  

In order to enlarge the dataset,
we have also searched the ASCA public archive for spectra of sources 
fulfilling our selection  criterium  and not yet available in the 
literature. Data preparation has been done using version 1.3 of the XSELECT
software package and version 4.0 of FTOOLS. Good time intervals were selected 
by applying the "Standard REV 2 Screening" criteria (as reported in Chapter 
5 of the ASCA Data Reduction Guide, rev 2.0). Only GIS2 and GIS3 data have been
analyzed; HIGH,MEDIUM and LOW bit rate data have been combined together.
Source counts were extracted from a circular region around the centroid of 
the X-ray emission whose radius maximize
the signal to noise ratio; similarly background data were taken from circular 
uncontaminated regions close to the X-ray source. 
The gis2v4-0.rmf and gis3v4-0.rmf
were used as detector redistribution matrix(RMF). 
The Ancillary Response Files (ARF) were created with the version 2.7
ASCAARF at the source location in the detector. In order to improve the 
statistics a combined GIS2+GIS3 spectrum was produced 
together with its respective background and response matrix files following
the recipe given in the ASCA Data Reduction (rev2.0) Guide. 
In order to use the ${\chi^2}$ statistics in the fitting procedure
spectral data have been rebinned to give at least 20 counts per bin. Only
for 3 sources  (NGC3281, NGC7590 and F17020+4544) 
useful data could be obtained (mostly due to the  
low signal to noise ratio of the observations); in 
these cases count rates have been 
analyzed by using the 
XSPEC 10.0 (X-ray Spectral Fitting) software package for data modelling.
Each of these 
observations have been fitted with a simple  model 
 consisting of an absorbed
power law plus a narrow gaussian line and an additional power law 
component to take into account the  soft X-ray emission; the spectral indices
of the two power law components have been tied together. 
Fitting the data with more complex models, as suggested by recent works (see 
for example \cite{turn97a} and references therein), is beyond the 
purpose of this work and is therefore postponed to a future paper.

\placetable{tbl-2} 

The number beside the name of each object indicates the reference 
(listed at the end in a footnote) from which the parameters reported 
in Table 2 were obtained; the letter beside some sources, those identified 
as Compton thick in section 2, indicate the lower limit available
on  N$_{\rm H}$.
For some sources more than one dataset was available in the literature:
in order to facilitate the use of the catalog only one entry, generally
that representing the most complete, recent or best fit publication 
has been included in Table 2. 
The list of references contained in Table 1 and 2 should represent to 
the best 
of our knowledge a complete survey of the literature up to the end of 1997,
giving the largest sample of hard
X-ray data on Seyfert 2 galaxies yet available.  
The complete catalog is available in computer readable form (ASCII file).
Copies can be requested from L. Bassani (Istituto TESRE/CNR-
Via Gobetti 101, 40129 Bologna, ITALY, loredana@botes1.tesre.bo.cnr.it).

\scriptsize

\begin{deluxetable}{rlllllll}
\tablewidth{0pt}
\tablecaption{Catalog of Seyfert 2. \label{tbl-1}}
\tablehead{
 &\colhead{Name}           
 &\colhead{RA \& Dec. (2000) }      
 &\colhead{z}          
 &\colhead{class$^{\star}$}  
 &\colhead{N$_{H,gal}^{\S}$}
 &\colhead{H$_{\alpha}$/H$_{\beta}$}
 &\colhead{F$_{OIII}^{\dagger}$}}
\startdata
1 &MKN 937        &00 10 10.0  $-$04 42 38  & 0.02952  &S2$^{a}$/S1$^{f}$ & 
3.34&  $-$         &     $-$  \nl

2&MKN348         &00 48 47.1  +31 57 25  & 0.01514  &  S2$^{a}$    &  5.86&
6.02$^{1}$   &  0.176$^{1}$  \nl

3&IC1631         &01 08 44.8  $-$46 28 33  &0.03084  &  S2$^{a}$    & 2.17&
7.30$^{2}$  & 0.052$^{2}$ \nl

4&NGC526a        &01 23 54.2  $-$35 03 56  & 0.01922  &  S2$^{b}$/S1.5$^{a}$&
2.20  & 3.00$^{3}$   &  0.027$^{3 }$   \nl

5&NGC1068        &02 42 40.6  $-$00 00 48  & 0.00379  & S2$^{c}$    & 3.53&
    7.00$^{4,5}$&   15.86${\pm2.7}$ $^{4,5}$\nl  

6&NGC1275        &03 19 48.1  +41 30 42  & 0.01756  &  S1.9$^{c}$/S1.5$^{g}$
&14.9 &    5.00$^{6}$& 0.311$^{6}$      \nl

7&NGC1365        &03 33 36.3  $-$36 08 26  & 0.00546  & S1.8$^{c}$    &1.39 &
   8.70 $^{5}$   & 0.141$^{5}$       \nl

8&NGC1386        &03 36 45.3  $-$35 59 57  & 0.00290  &  S2$^{c}$    &1.37  & 
  5.70 $^{7,8}$  &   0.655${\pm0.24}$ $^{7,8}$\nl

9&MCG$-$01$-$12$-$006  &04 25 55.6  $-$08 34 07  & 0.03920  &  S2$^{a}$ &6.23 &
     11.50$^{2}$  &   0.018$^{2}$ \nl

10&NGC1672        &04 45 42.1  $-$59 14 57  & 0.00450  &  S2$^{a}$    &2.28   &
6.80$^{4,5}$  & 0.077 ${\pm0.03}$ $^{4,5}$   \nl

11&NGC1667        &04 48 37.1  $-$06 19 12  & 0.01517  &  S2$^{c}$    
&5.49   &9.74$^{4}$  & 0.197$^{4}$    \nl

12&[HB89]0449$-$183 &04 51 37.3  $-$18 18 44  & 0.33800  &  S2$^{a}$    &3.88  & 
 7.00$^{9}$         & 0.005$^{9}$   \nl

13&F04575$-$7537    &04 55 59.6  $-$75 32 27  & 0.01810  &  S2$^{a}$    &8.56   &
12.5$^{10}$    & 0.331$^{10}$      \nl

14&NGC1808        &05 07 42.3  $-$37 30 46  & 0.00334  &  S2$^{a}$/SB$^{i}$    & 2.70  &
14.2$^{5}$ & 0.131$^{5}$   \nl

15&F05189$-$2524    &05 21 01.4  $-$25 21 45  & 0.04256  & S2$^{a}$    &1.92 &
 23.4$^{11}$       & 0.083$^{11}$        \nl

16&NGC2110        &05 52 11.4  $-$07 27 22  & 0.00762  &  S2$^{c}$    &18.3   &

8.15$^{1}$  & 0.321$^{1}$     \nl

17&MKN3           &06 15 36.3  +71 02 15  & 0.01351  &  S2$^{a}$    &8.38   &
6.67$^{1}$  & 4.600$^{1}$ \nl

18&NGC2273        &06 50 08.7  +60 50 45  & 0.00614  &  S2$^{c}$    &6.97   &
  6.92$^{12}$   & 0.277${\pm0.08}$$^{12,13}$    \nl

19&MKN1210        &08 04 05.9  +05 06 50  & 0.01350  &  S2$^{a}$    & 3.73  &
5.20$^{14}$       & 0.482$^{14}$      \nl

20&NGC2639        &08 43 38.0  +50 12 20  & 0.01062  &  S1.9$^{c,g}$&3.00   &
4.16$^{6}$       &$>$0.005$^{6}$ \nl

21&F09104+4109    &09 13 44.0  +40 56 34  & 0.44200  &  S2$^{a}$    &0.98  &
 3.00 $^{15}$      & 0.035${\pm0.020}$  $^{15,8}$        \nl

22&NGC2992        &09 45 42.0  $-$14 19 35  & 0.00771  &  S1.9$^{c}$    &5.26 &  
6.97$^{1}$  &    0.680${\pm0.027}$$^{1,13}$ \nl

23&MCG$-$5$-$23$-$16    &09 47 40.2  $-$30 56 54  & 0.00828  &  S2$^{a}$ &8.00
&   8.00$^{2}$  & 0.409$^{2}$   \nl

24&NGC3031/M81    &09 55 33.2  +69 03 55  & $-$0.0001 &S1.8$^{c}$/S1.5$^{g,f}$
&4.16 & 3.19$^{16}$           &0.043$^{16}$        \nl

25&NGC3081        &09 59 29.5  $-$22 49 35  & 0.00796  &  S2$^{c}$    &4.61  & 
4.40$^{4}$       & 0.215${\pm0.02}$ $^{4,17}$        \nl

26&NGC3079        &10 01 57.8  +55 40 47  & 0.00375  &  
S2$^{a}$/L$^{f}$/SB$^{i}$    
&7.89   &25.0$^{6}$     & 0.090$^{6}$        \nl

27&NGC3147        &10 16 53.6  +73 24 03  & 0.00941  &  S2$^{a}$    &3.64  &
5.26$^{6}$           & 0.009$^{6}$        \nl

28&NGC3281        &10 31 52.0  $-$34 51 12  & 0.01154  &  S2$^{c}$    &6.42  &
 6.13$^{8}$ &   0.045$^{13}$\nl 

29&MCG+12$-$10$-$067  &10 44 08.7  +70 24 19  & 0.03280  & S2$^{a}$    &2.64 &  
7.20$^{2}$  &0.075$^{2}$    \nl

30&NGC3393        &10 48 24.0  $-$25 09 40  & 0.01370  &  S2$^{c}$    &6.05   &
4.12$^{18}$     &0.316$^{18}$   \nl

31&NGC4258        &12 18 57.5  +47 18 14  & 0.00149  &  S1.9$^{c}$    & 1.16  &
 9.12$^{19}$    & 0.262$^{19}$      \nl

32&NGC4388        &12 25 47.2  +12 39 40  & 0.00842  &  S2$^{c}$/SB$^{i}$    &2.60   &
5.50$^{1,5}$  & 0.374${\pm0.050}$ $^{1,5}$ \nl

33&NGC4507        &12 35 36.5  $-$39 54 33  & 0.01180  &  S2$^{c}$    &7.23  & 
4.50$^{1,20}$  & 0.158${\pm0.060}$ $^{1,20}$   \nl

34&NGC4579        &12 37 43.5  +11 49 05  & 0.00507  &  S1.9$^{c}$/L$^{g}$    &2.47  &
3.22$^{6}$           &  0.009$^{6}$         \nl

35&NGC4594        &12 39 58.8  $-$11 37 28  & 0.00364  & S1.9$^{c}$/L$^{g,f}$ 
   &3.77 &3.45$^{6}$            & 0.007$^{6}$         \nl

36&NGC4939        &13 04 43.3  $-$10 20 23  & 0.01038  &  S2$^{c}$    &3.35 &  
  4.50$^{5}$     & 0.112$^{5}$  \nl

37&NGC4941        &13 04 13.1  $-$05 33 06  & 0.00370  &  S2$^{c}$    &2.42  & 
  6.80$^{21,7}$     & 0.355$\pm0.090$ $^{21,7}$       \nl

38&NGC4945        &13 05 27.5  $-$49 28 06  & 0.00187  &  S2$^{c}$/SB$^{i}$    &15.7   &
$-$        &  $-$      \nl

39&NGC4968        &13 07 06.0  $-$23 40 43  & 0.00986  &  S2$^{a}$    & 9.16  &
14.9$^{22}$  & 1.116$^{22}$  \nl

40&NGC5033        &13 13 27.3 $+$36 35 36   &0.00292   & S1.9$^{c}$/S1.5$^{g}$  & 1.01&
4.48$^{6}$ &       0.017  $^{6}$   \nl

41&AXJ131501$+$3141 &13 15 01.1 +31 41 28 & 0.07200  & S2$^{d}$  & 1.11 &
7.58$^{23}$  & $-$ \nl

42&PKS B1319$-$164  &13 22 24.5  $-$16 43 43  & 0.01718  &  S1.8$^{a}$ & 5.81  &   
4.17$^{24}$       & 0.401$^{24}$   \nl

43&NGC5128/CENA   &13 25 27.6  $-$43 01 08  & 0.00183  &  S2$^{c}$   &8.62  &  
  5.50$^{5}$    &0.007$^{5}$ \tablenotemark{N}    \nl

44&NGC5135        &13 25 49.9  $-$29 50 02  & 0.01372  &  S2$^{c}$    &4.65 &  
7.80$^{7}$       & 0.614$^{7}$ \nl

45&NGC5194/M51        &13 29 52.3  +47 11 54  & 0.00154  &  S2$^{c}$    & 1.57 & 
8.33$^{6}$           &0.228$^{6}$             \nl

46&NGC5252        &13 38 15.9  +04 32 33  & 0.02298  &  S1.9$^{a}$    & 1.97  &
3.72$^{25}$  & 0.039$^{17}$  \nl

47&MKN273         &13 44 42.1  +55 53 13  & 0.02778  &  S2/L$^{a}$  &1.09   &
9.33$^{1}$  & 0.084$^{1}$  \nl

48&MKN463E        &13 56 02.6  +18 22 18  & 0.05100  &  S2$^{a}$    &2.06   &
5.59$^{1}$  & 0.124$^{1}$  \nl

49&CIRCINUS       &14 13 10.2  $-$65 20 21  & 0.00145  &  S2$^{c}$/SB$^{i}$    &55.6  & 
19.1$^{26}$ &  6.970$^{27}$  \nl

50&NGC5506        &14 13 14.8  $-$03 12 27  & 0.00618  &  S1.9$^{c}$    &3.81   &
7.20$^{1,4}$  &0.600${\pm0.010}$$^{1,4}$    \nl

51&NGC5643        &14 32 40.7  $-$44 10 28  & 0.00400  &  S2$^{c}$    &8.34   &
6.40$^{1,4,5}$  & 0.694${\pm0.08}$ $^{1,4,5}$   \nl

52&NGC5674        &14 33 52.2  +05 27 30  & 0.02492  &  S1.9$^{c}$    &2.46   &
4.90$^{25}$  & 0.059$^{17}$  \nl

53&MKN477         &14 40 38.1  +53 30 16  & 0.03780  &  S2$^{b}$/NS1$^{h}$    &1.30   &
5.40$^{1}$     & 1.238$^{1}$  \nl

54&NGC6251        &16 32 31.8  +82 32 16  & 0.02302  &  S2$^{a}$    &5.49   &
15.1$^{28}$ &  0.057$^{28}$ \nl

55&NGC6240  &16 50 27.5  +02 28 58  & 0.02448  &  S2-L$^{a}$/SB$^{i}$  & 5.68  &
17.2$^{11,29,30}$    & 0.135${\pm0.02}$ $^{11,29,30}$      \nl

56&F17020+4544    &17 03 30.3  +45 40 46  & 0.06040  &  S2$^{a}$ /NS1$^{f}$   &2.22   &
5.40$^{31}$ & 0.024$^{31}$         \nl

57&MKN507         &17 48 38.4  +68 42 16  & 0.05590  &  S2$^{a}$/NS1$^{f}$    &4.37 &        
  4.70$^{1}$  & 0.002$^{1}$         \nl

58&AXJ1749+684    &17 49 49.0  $^{b}$68 23 03  & 0.05000  &  NLXG$^{e}$ & 4.48 &   
 7.32$^{32}$ & 0.004$^{32}$ \nl

59&NGC6552        &18 00 07.2  +66 36 55  & 0.02600  &  S2$^{a}$    &4.23   &
5.11$^{33}$  & 0.096$^{33}$   \nl

60&F18325$-$5926    &18 36 57.9  $-$59 24 09  & 0.02023  &  S1.8$^{c}$    &7.15  &   
10.6$^{31,34}$  & 0.752${\pm0.015}$$^{31,34}$    \nl

61&ESO103$-$G35     &18 38 20.1  $-$65 25 42  & 0.01329  &  S2$^{a}$    & 7.64  &
6.31$^{2}$  & 0.112$^{2}$    \nl

62&CYGNUS A       &19 59 28.2  +40 44 02  & 0.05605  &  S2$^{b}$    & 34.7  &
 5.40$^{35}$     & 0.080$^{35}$  \nl

63&F20210+1121    &20 23 25.9  +11 31 31  & 0.05639  &  S2$^{a}$    &13.8  &
6.40$^{36,37}$  & 0.314${\pm0.05}$$^{36,37}$        \nl

64&IC5063         &20 52 01.9  $-$57 04 09  & 0.01135  &  S2$^{c}$    &6.73 &  
5.80$^{38,8}$  &   0.353${\pm0.190}$$^{38,8}$ \nl

65&F20460+1925    &20 48 17.9  +19 36 57  & 0.18100  &  S2$^{a}$    & 11.2  &
7.08$^{37,39}$     & 0.103$^{37,39}$  \nl

66&IC 1368       &21 14 12.1   +02 10 38    & 0.01305 &  S2$^{a}$    &6.55 &
   $-$            &   $-$             \nl

67&NGC7172        &22 02 01.6  $-$31 52 19  & 0.00868  &  S2$^{c}$    &1.65 &  
3.00$^{40}$ &    0.004$^{40}$   \nl

68&NGC7314        &22 35 46.0  $-$26 03 03  & 0.00474  &  S1.9$^{c}$    &1.46  & 
20.0$^{5}$  & 1.770$^{5}$  \nl

69&NGC7319        &22 36 03.4  +33 58 33  & 0.02256  &  S2$^{a}$    & 8.01  &
4.72$^{41}$       & 0.024$^{41}$ \nl

70&F23060+0505    &23 08 34.0  +05 21 30  & 0.17300  &  S2$^{a}$    &5.51 &
$>$13.8$^{31}$           & $>$0.144$^{31}$         \nl

71&NGC7582        &23 18 23.4  $-$42 22 15  & 0.00525  &  S2$^{c}$    & 1.93 &   
7.60$^{4,5}$  & 0.445${\pm0.045}$$^{4,5}$   \nl

72&NGC7590        &23 18 55.0  $-$42 14 17  & 0.00532  &  S2$^{c}$    &1.96 &
5.40$^{4}$            & 0.017$^{4}$   \nl

73&NGC7674        &23 27 56.6  +08 46 44  & 0.02906  &  S2$^{a}$    &5.15   &
4.80$^{1,28}$  & 0.185${\pm0.01}$$^{1,28}$  \nl

\tablenotetext{\star}{Classification is as follows S=Seyfert, L=Liner, SB=Starburst}
\tablenotetext{\S}{Galactic HI column density from 21 cm measurements 
in units of 10$^{20}$ cm$^{-2}$ from HEASARC on line service}
\tablenotetext{\dagger}{Corrected [OIII] fluxes in units of 10$^{-11}$
erg cm$^{-2}$ s$^{-1}$}
\tablenotetext{N}{For this object the only [OIII] measurement available 
was taken with an aperture smaller than 100pc as projected on the source 
and, therefore, 
this data are likely to miss a significant fraction of the [OIII] flux.}  

\tablerefs{
($^{a}$)NED;
($^{b}$)SIMBAD;
($^{c}$)\cite{mari95};
($^{d}$)\cite{akiy98};
($^{e}$)\cite{iwas97a};
($^{f}$)\cite{mora96};
($^{g}$)\cite{ho97};
($^{h}$)\cite{vero97};
($^{i}$)\cite{lh96}
}

\tablerefs{
($^{1}$)\cite{dade88};
($^{2}$)\cite{poal96};
($^{3}$)\cite{wink92};
($^{4}$)\cite{stbe95};
($^{5}$)\cite{vece86};
($^{6}$)\cite{ho97};
($^{7}$)\cite{stbe89};
($^{8}$)\cite{phil83};
($^{9}$)\cite{step89};
($^{10}$)\cite{vign98};
($^{11}$)\cite{veil95};
($^{12}$)\cite{lons92};
($^{13}$)\cite{witt92};
($^{14}$)\cite{terl91};
($^{15}$)\cite{craw96};
($^{16}$)\cite{ho93};
($^{17}$)\cite{cruz94};
($^{18}$)\cite{diaz88};
($^{19}$)\cite{heck80};
($^{20}$)\cite{durr86};
($^{21}$)\cite{stau82};
($^{22}$)\cite{oste85};
($^{23}$)\cite{akiy98};
($^{24}$)\cite{dero88};
($^{25}$)\cite{oste93};
($^{26}$)\cite{oliv94};
($^{27}$)Oliva, 1998 private communications;
($^{28}$)\cite{shud81};
($^{29}$)\cite{kim95};
($^{30}$)\cite{laur89};
($^{31}$)\cite{degr92};
($^{32}$)\cite{iwas97a};
($^{33}$)Moran, 1997 private communication;
($^{34}$)\cite{iwas95};
($^{35}$)\cite{stoc94};
($^{36}$)\cite{pere90};
($^{37}$)\cite{vade93};
($^{38}$)\cite{berg83};
($^{39}$)\cite{frog89};
($^{40}$)\cite{vace97};
($^{41}$)\cite{keel85}.}
\enddata
\end{deluxetable}

\newpage

\begin{deluxetable}{rlccccc}
\scriptsize
\tablewidth{0pt}
\tablecaption{Seyfert 2 spectral parameters. \label{tbl-2}}
\tablehead{
 &\colhead{Name}           
 &\colhead{$\Gamma$}      
 &\colhead{N$_{H}^{\S}$}          
 &\colhead{EW(Fe$_{k \alpha}^{\star}$)}  
 &\colhead{F$_{x,obs.}^{\dagger}$}
 &\colhead{F$_{x,corr.}^{\dagger}$}}
 \startdata
1&MKN937$^{1}$& 1.99$^{+0.06}_{-0.04}$ & 3.3$^{+2.7}_{-0}$\tablenotemark{pl} &
 143$^{+207}_{-143}$  & 0.19   & 0.19  \nl
& & & & & & \nl

2 & MKN348$^{2}$     & 1.59$^{+0.15}_{-0.14}$ & 1060$^{+310}_{-260}$    &  
 230$^{+120}_{-140}$ & 1.27   & 2.21  \nl
& & & & & & \nl

3 & IC1631$^{3}$     & 2.10$^{+0.10}_{-0.10}$ &      $<$ 31.6         &
 $<$ 70               & 1.00  &1.00  \nl
& & & & & & \nl   

4 & NGC526A$^{1}$    & 1.82$^{+0.16}_{-0.14}$ & 150$^{+14}_{-14}$ &  
  111$^{+33}_{-56}$   & 3.44   & 3.69  \nl
& & & & & & \nl

5 & NGC1068$^{4,A}$   & 0.35$^{+0.66}_{-0.35}$   & 4.5                   & 
  1210$^{+260}_{-280}$& 0.35   & 0.35 \nl
& & & & & & \nl

6 & NGC1275$^{5}$   & 2.65$^{+0.20}_{-0.40}$   & 149$^{+69}_{-69}$&
 -                    & 9.55   & 11.4 \nl
& & & & & & \nl

7 & NGC1365$^{6}$   & 1.80$^{+0.14}_{-0.13}$      & 2000$^{+400}_{-400}$ & 
 320$^{+90}_{-160}$   &  0.53   & 1.03 \nl

& & & & & & \nl

8 & NGC1386$^{7,A}$ & 1.70$^{f}$ &  1.4          & 
 7600$^{+8900}_{-5000}$            &  0.02  & 0.02  \nl
& & & & & & \nl

9 & MCG$-$01$-$12$-$006$^{8}$&   1.95$^{+0.36}_{-0.40}$   &    106$^{+69}_{-79}$ &
 -              &7.30    & 8.06 \nl
& & & & & & \nl

10 & NGC1672$^{9}$    & 1.50$^{+0.20}_{-0.20}$   &       63$^{+253}_{-63}$  & 
 $<$150                 & 0.30 & 0.37 \nl
& & & & & & \nl

11& NGC1667$^{1,B}$     &   3.44$^{+1.31}_{-0.44}$  &   
5.5$^{+25.3}_{-0}$\tablenotemark{pl} &
 $<$3000                & 2.6e$-$3  &  2.6e$-$3  \nl
& & & & & & \nl

12&[HB89]E0449$-$184$^{1}$    &   4.12$^{+0.22}_{-0.63}$   & 17$^{+17}_{-10}$&
 61$^{+971}_{-61}$     & 0.08 & 0.10  \nl
& & & & & & \nl

13& F04575$-$7537$^{10}$  & 1.49$^{+0.07}_{-0.06}$         & 105$^{+10}_{-10}$  &
 142$^{+49}_{-50}$     & 1.20  & 1.36 \nl
& & & & & & \nl

14& NGC1808$^{1}$ &1.49$^{+1.15}_{-1.32}$ &320$^{+588}_{-318}$\tablenotemark{l}  &
 43$^{+586}_{-43}$     & 0.09 & 0.11 \nl
& & & & & & \nl

15 & F05189$-$2524$^{11}$ &  1.70$^{f}$         & 490$^{+10}_{-16}$  &
 $<$ 30      &0.43  &0.61  \nl
& & & & & & \nl

16 & NGC2110$^{12}$    & 1.36$^{+0.07}_{-0.08}$      & 289$^{+21}_{-29}$   & 
 124$^{+36}_{-36}$     & 2.60  & 3.20  \nl
& & & & & & \nl

17  & MKN3$^{13}$    &1.56$^{+0.14}_{-0.26}$        & 11000$^{+1500}_{-2500}$& 
997$^{+300}_{-307}$   & 0.65  & 4.08 \nl
& & & & & & \nl

18  & NGC2273$^{7,A}$  &0.70$^{+0.5}_{-0.5}$      &$<$300          & 
 3800$^{+1100}_{-1100}$& 0.09 & 0.09  \nl
& & & & & & \nl

19 & MKN1210$^{14,B}$ &0.90 $^{+1.00}_{-1.10}$    &1200$^{+1100}_{-1200}$ &
820$^{+360}_{-430}$   & 0.13 & 0.21 \nl
& & & & & & \nl

20  & NGC2639$^{15}$ & 2.4$^{+0.4}_{-0.5}$   &4200$^{+5600}_{-2300}$ &
-                      & 0.04 & 0.2  \nl
& & & & & & \nl

21 & F09104+4109$^{16,cf}$    &1.96$^{+0.06}_{-0.06}$    
&24$^{+6}_{-6}$ &
308$^{+90}_{-120}$$^{i}$   & 0.12 & 0.12  \nl
& & & & & & \nl

22 & NGC2992$^{17}$   & 1.70\tablenotemark{f}        &  69$^{+33}_{-19}$      &
514$^{+190}_{-190}$   & 0.45  & 0.45  \nl
& & & & & & \nl

23 &MCG$-$5$-$23$-$16\tablenotemark{18}  &   1.95$^{+0.10}_{-0.09}$      & 162$^{+23}_{-21}$    & 
 62$^{+31}_{-24}$      & 7.30  & 8.62 \nl
& & & & & & \nl

24 & NGC3031/M81$^{19}$ &   1.85$^{+0.04}_{-0.04}$   & 9.4$^{+0.7}_{-0.6}$    &
170$^{+60}_{-60 }$$^{i}$       & 1.50 & 1.50  \nl
& & & & & & \nl

25 & NGC3081$^{7}$&  1.70$^{+0.26}_{-0.35}$     & 6600$^{+1800}_{-1600}$    &
 610$^{+390}_{-210}$  &  0.13  & 0.68  \nl
& & & & & & \nl

26 & NGC3079$^{20}$  &   1.76$^{+1.02}_{-0.89}$      & 160$^{+270}_{-130}$  &
 -                    & 0.053  &0.06  \nl
& & & & & & \nl

26 & NGC3147$^{7}$    &   1.80$^{+0.09}_{-0.09}$  & 4.3$^{+3.2}_{-2.7}$  &
485$^{+309}_{-282}$       & 0.13  & 0.13 \nl
& & & & & & \nl

28 & NGC3281$^{22}$  &  1.47$^{+0.23}_{-0.25}$      &7980$^{+1900}_{-1500}$  &
 751$^{+450}_{-267}$ & 0.40  & 2.84 \nl
& & & & & & \nl

29& MCG+12$-$10$-$067$^{9}$ &  1.40$^{+0.06}_{-0.06}$   &1479$^{+516}_{-382}$    &
 $<$240               & 0.30 & 0.30  \nl
& & & & & & \nl

30 & NGC3393$^{7,A}$  &  -035$^{+0.50}_{-0.24}$       & $<$ 7000      &
 3500$^{+2000}_{-2000}$ & 0.04 & 0.04 \nl
& & & & & & \nl

31& NGC4258$^{23}$   &   1.78$^{+0.29}_{-0.29}$     & 1500$^{+200}_{-200}$ & 
 250$^{+100}_{-100}$   & 0.30 & 0.64 \nl
& & & & & & \nl

32& NGC4388$^{24}$   &   1.60$^{+0.50}_{-0.40}$     & 4200$^{+600}_{-1000}$ &
 732$^{+243}_{-191}$   & 1.20 & 4.30 \nl
& & & & & & \nl

33 & NGC4507$^{25}$ &    1.61$^{+0.20}_{-0.20}$    & 2920$^{+230}_{-230}$ & 
189$^{+36}_{-36}$     & 2.10  & 7.03 \nl
& & &  & & & \nl

34 & NGC4579$^{26}$ &   1.72$^{+0.05}_{-0.05}$       & 4.1$^{+2.7}_{-2.7}$ &
240$^{+170}_{-160}$       & 0.44  & 0.44 \nl
& & &  & & & \nl

35 & NGC4594$^{20}$ &  1.96$^{+0.21}_{-0.20}$      & 55$^{+31}_{-40}$ &
 - & 0.19  & 0.195  \nl
& & &  & & & \nl

36 & NGC4939$^{7}$    & 1.70\tablenotemark{f}                 & 3000$^{+2900}_{-1800}$ &
  490$^{+410}_{-290}$    & 0.15  & 0.45  \nl
& & & & & & \nl

37  & NGC4941$^{7}$  & 1.70\tablenotemark{f}      &4500$^{+2500}_{-1400}$& 
 1600$^{+700}_{-900}$   & 0.07 & 0.30  \nl
& & & & & & \nl

38  & NGC4945$^{27}$ & 1.82$^{+0.08}_{-0.09}$ & 40000$^{+2000}_{-1200}$&  
 850$^{+160}_{-160}$  & 0.35  & 10.2 \nl
& & & & & & \nl

39  & NGC4968$^{1,B}$    & 1.33$^{+3.07}_{-1.33}$      
&8.4$^{+4.6}_{-0}$\tablenotemark{pl} &  
 1180$^{+4420}_{-827}$ & 0.04  & 0.04 \nl
& & & & & & \nl

40 & NGC5033$^{28}$    & 1.72$^{+0.04}_{-0.04}$ & 8.7$^{+1.7}_{-1.7}$&
290$^{+100}_{-100}$ &0.55  & 0.55 \nl
& & & & & & \nl

41 &AXJ131501+3141$^{29}$&  1.50$^{+0.70}_{-0.60}$  & 600$^{+400}_{-200}$ &
 $-$   & 0.05 & 0.07 \nl
& & & & & & \nl

42 &PKS B1319$-$164$^{14}$&  4.10$^{+0.80}_{-0.80}$  &7600$^{+1300}_{-1220}$ &
 200$^{+110}_{-100}$   & 0.21 & 0.65 \nl
& & & & & & \nl

43 & NGC5128/CenA$^{30}$        & 1.96$^{+0.10}_{-0.10}$   &  1000-3500          & 
 114$^{+18}_{-18}$     & 8.50 & 20.4  \nl
& & & & & &  \nl

44 & NGC5135$^{1,B}$  &   2.91$^{+0.33}_{-0.15}$      
& 4.7$^{+7.6}_{-0}$\tablenotemark{pl}  & 
 11700$^{+2631}_{-11605}$& 0.02 & 0.02  \nl
& & & & & & \nl

45 & NGC5194/M51$^{31}$ &   1.43\tablenotemark{f} &5000-10000  &
1100$^{+600}_{-600}$       & 0.11 & 0.14-0.36  \nl
& & & & & & \nl

46& NGC5252$^{32}$   &  1.45$^{+0.20}_{-0.20}$      & 433$^{+66}_{-61}$  & 
 78$^{+44}_{-53}$      & 0.89  &  1.16 \nl
& & & & & & \nl

47 & MKN273$^{1}$    &  2.25$^{+2.75}_{-4.14}$\tablenotemark{ph}   & 4902$^{+6697}_{-4903}$ &
 342$^{+236}_{-342}$   & 0.06  & 0.35 \nl
& & & & & & \nl

48 & MKN463E$^{33}$   & 1.40$^{+0.90}_{-0.90}$     & 1600$^{+800}_{-800}$ &
    $<$ 670        & 0.04  & 0.09 \nl
& & & & & & \nl

49  & CIRCINUS$^{34,B}$   &  1.60$^{+0.50}_{-0.40}$      &23$^{+9}_{-9}$ &
 2310$^{+120}_{-260}$  & 1.00   & 1.00 \nl
& & & & & &  \nl

50 & NGC5506$^{2}$    &  1.92$^{+0.03}_{-0.02}$      & 340$^{+26}_{-12}$ &
   150$^{+30}_{-30}$                & 8.38 & 10.8  \nl
& & & & & & \nl

51 & NGC5643$^{7,A}$    & 1.53$^{+0.23}_{-0.26}$            & $<$23          &
 1800$^{+800}_{-960}$  & 0.13  & 0.13  \nl
& & & & & & \nl

52 & NGC5674$^{2}$    & 1.66$^{+0.17}_{-0.14}$          & 700$^{+280}_{-260}$ &
     $<$ 140           & 0.84   & 1.28 \nl
& & & & & &  \nl

53 &MKN477$^{14,B}$      & 0.2$^{+0.80}_{-0.70}$     & 900$^{+1200}_{-900}$ &
 490$^{+250}_{-200}$   & 0.12 & 0.16 \nl
& & & & & & \nl

54 & NGC6251$^{1}$    &  2.14$^{+0.09}_{-0.09}$      & 12.4$^{+3.6}_{-3.4}$
& 392$^{+306}_{-305}$\tablenotemark{i} & 0.13 &0.13 \nl
& & & & & & \nl

55 & NGC6240$^{35,B}$   &   0.30$^{+0.60}_{-0.50}$    &130$^{+180}_{-130}$&   
1580$^{+380}_{-350}$  & 0.19   & 0.19 \nl
& & & & & & \nl

56&F17020+4544$^{22}$    &  2.04$^{+0.07}_{-0.03}$          &  $<$ 2.8 &
 168  $^{+140}_{-80}$ &0.53  &0.53 \nl
& & & & & & \nl
                           
57&MKN507$^{36}$    &1.76 $^{+0.22}_{-0.20}$               &34$^{+15}_{-13}$ &
 $<$ 730         &0.05& 0.05 \nl 
& & & & & & \nl

58&AXJ1749+684$^{37}$    &   1.40$^{+0.55}_{-0.37}$          & 21$^{+62}_{-21}$ &
  $<$ 1700     &0.96  & 0.96 \nl
& & & & & & \nl

59&NGC6552$^{38}$    &   1.40$^{+0.4}_{-0.4}$ &      6000$^{+1000}_{-1000}$ &
 900               & 0.06 & 0.21 \nl
& & & & & & \nl

60&F18325$-$5926$^{39}$&    2.32$^{+0.06}_{-0.06}$    & 132$^{+10}_{-10}$   & 
 580$^{+400}_{-250}$   & 0.84 & 1.01 \nl
 & & & & & &\nl

61 & ESO103$-$G35$^{1}$ & 1.52$^{+0.34}_{-0.06}$    & 1591$^{+10410}_{-1262}$ & 
 173$^{+51}_{-116}$    & 0.90 & 1.83 \nl
& & & & & & \nl

62& CYGNUS A$^{40,cf}$    &     1.98$^{+0.18}_{-0.20}$   & 3750$^{+750}_{-710}$      & 
 380$^{+80}_{-80}$\tablenotemark{i}   & 1.80 & 7.79 \nl
& & & & & &  \nl

63& F20210+1121$^{41,A}$    &  0.40$^{+0.90}_{-0.90}$   & $<$ 600  &
 1600\tablenotemark{i}    & 0.03 & 0.03 \nl 
& & & & & &  \nl

64 & IC5063$^{14}$  & 1.80$^{+0.20}_{-0.2}$      &2400$^{+200}_{-200}$     & 
 80$^{+42}_{-50}$      & 1.20 & 3.0 \nl
& & & & & & \nl

65 &F20460+1925$^{42}$   &  1.99$^{+0.18}_{-0.15}$     &250$^{+34}_{-32}$    & 
 260$^{+145}_{-137}$   & 0.15& 0.19 \nl
& & & & & &  \nl

66& IC1368$^{3}$    &  2.10$^{+0.10}_{-0.10}$   & $<$ 159  &
 $<$ 180  & 1.30 & 1.30 \nl
& & & & & &  \nl

67 & NGC7172$^{43}$   &  1.52$^{+0.14}_{-0.15}$    & 861$^{+79}_{-33}$      & 
 121$^{+50}_{-60}$    & 1.30 & 2.14 \nl
& & & & & &  \nl

68 & NGC7314$^{1}$     & 2.41$^{+0.05}_{-0.02}$    &116$^{+4}_{-13}$    & 
 182$^{+46}_{-47}$     & 3.56 & 4.05 \nl
& & & & & & \nl

69 &NGC7319$^{14}$  &  1.40$^{+0.8}_{-1.6}$    &3300$^{+1400}_{-2200}$    & 
 620$^{+230}_{-260}$   & 0.12 & 0.33 \nl
& & & & & & \nl

70 &F2306+0505$^{44}$  & 2.14 $^{+0.66}_{-0.53}$    &840$^{+190}_{-250}$    &
 170$^{+120}_{-170}$  &0.15  &0.25  \nl
& & & & & & \nl

71 & NGC7582$^{45}$    &1.52$^{+0.09}_{-0.07}$     &1240$^{+60}_{-80}$      & 
 182$^{+50}_{-40}$     & 1.55  & 2.72 \nl
& & & & & & \nl

72 & NGC7590$^{22}$    & 2.29$^{+0.20}_{-0.13}$        &$<$ 9.2      &
  $-$     &0.12  & 0.12  \nl
& & & & & & \nl

73 & NGC7674$^{46,A}$    &  1.92 $^{+0.21}_{-0.21}$             & 5.3 &
 900$^{+470}_{-299}$ & 0.05 &  0.05 \nl

\tablenotetext{\star}{Fe$_{k\alpha}$ line equivalent width in units of eV}
\tablenotetext{\S}{absorbing column density in the source direction 
in units of 10$^{20}$ cm$^{-2}$} 
\tablenotetext{\dagger}{2 - 10 keV observed and absorption corrected 
fluxes in units of 10$^{-11}$ erg sec$^{-1}$ cm$^{-2}$}
\tablenotetext{f}{fixed photon index}
\tablenotetext{pl}{indicates the parameter is pegged at the lower limit}
\tablenotetext{ph}{indicates the parameter is pegged at the higher limit}
\tablenotetext{i}{line at 6.7 keV;}
\tablenotetext{cf}{X-ray emission maybe dominated by a cooling flow 
around F09104+4109 (\cite{fabi95}) and by intracluster emission in Cygnus A
\cite{ueno94})} 
\tablenotetext{A} {Compton thick source with N$_{\rm H}$ $\ge$ 10$^{25}$ 
cm$^{-2}$}
\tablenotetext{B} {Compton thick source with N$_{\rm H}$ $\ge$ 10$^{24}$
cm$^{-2}$}
\tablerefs{
($^{1}$)\cite{turn97a};  
($^{2}$)\cite{smdo96};
($^{3}$)\cite{awak92};
($^{4}$)\cite{iwas97c}; 
($^{5}$)\cite{kowa93}  ;
($^{6}$)Maiolino 1998, private comunication;
($^{7}$)\cite{maio98};
($^{8}$)\cite{poll96};
($^{9}$)\cite{awko93}; 
($^{10}$)\cite{vign98};
($^{11}$)\cite{kii96};
($^{12}$)\cite{haya96};
($^{13}$)\cite{capp98}; 
($^{14}$)\cite{ueno97};
($^{15}$)\cite{wils98};
($^{16}$)\cite{fabi94};
($^{17}$)\cite{weav96}; 
($^{18}$)\cite{weav97};
($^{19}$)\cite{ishi96};
($^{20}$)\cite{serl96};
($^{21}$)\cite{ptak96};
($^{22}$)this work;
($^{23}$)\cite{maki94};
($^{24}$)\cite{iwas97a};    
($^{25}$)\cite{coma98};
($^{26}$)\cite{ter98a};
($^{27}$)\cite{done96};
($^{28}$)\cite{ter98b};
($^{29}$)\cite{akiy98};
($^{30}$)\cite{turn97c};
($^{31}$)\cite{ter98c};
($^{32}$)\cite{capp96};
($^{33}$)\cite{ueno96};
($^{34}$)\cite{matt96b};
($^{35}$)\cite{iwco98};
($^{36}$)\cite{iwas98};
($^{37}$)\cite{iwas97b};
($^{38}$)\cite{fuka94};
($^{39}$)\cite{iwas96};
($^{40}$)\cite{ueno94};
($^{41}$)\cite{ueno98};
($^{42}$)\cite{ogas97};
($^{43}$)\cite{guai98};
($^{44}$)\cite{bran97};
($^{45}$)\cite{xue98};
($^{46}$)\cite{mala98}.}
\enddata
\end{deluxetable}

\end{document}